\documentclass[prl,a4paper,showpacs,twocolumn,nofootinbib,preprintnumbers]{revtex4}

\usepackage{graphicx}
\usepackage{dcolumn}
\usepackage{bm}
\usepackage{epsf}
\usepackage{dcolumn}
\usepackage{bm}
\def\be{\begin{equation}}
\def\ee{\end{equation}}
\def\bea{\begin{eqnarray}}
\def\eea{\end{eqnarray}}

\def\gsim{\ \rlap{\raise 2pt\hbox{$>$}}{\lower 2pt \hbox{$\sim$}}\ }
\def\lsim{\ \rlap{\raise 2pt\hbox{$<$}}{\lower 2pt \hbox{$\sim$}}\ }
\def\dslash{\kern-4pt \not{\hbox{\kern-2pt $\partial$}}}
\def\pslash{\not{\hbox{\kern-2pt p}}}

\def\petau{{${P_{e \tau}}$}}

\def\pemu{{${P_{e \mu }}$ }}

\def\pee{{${P_{ee}}$ }}

\newcommand{\nue}{\nu_e}
\newcommand{\anue}{\bar\nu_e}

\newcommand{\ma}{\Delta m^2_{31}}

\newcommand{\stch}{\sin^2 2\theta_{13}}

\newcommand{\mat}{\Delta m^2_{31}{\mbox {(true)}}}

\newcommand{\bb}{$\beta$-beam}

\newcommand{\chr}{\mbox{$\breve{\rm C}$erenkov~}}

\begin{document}


\title{Neutrino parameters from matter effects in $P_{ee}$ at
long baselines} 


\author{Sanjib Kumar Agarwalla$^{1,2}$}
\author{Sandhya Choubey$^{1}$}
\author{Srubabati Goswami$^{1}$}
\author{Amitava Raychaudhuri$^{1,2}$}

\affiliation{\vskip 0.1in
$^1$Harish-Chandra Research Institute, Chhatnag Road, Jhunsi,
Allahabad 211 019, India
\\
$^2$Department of Physics, University of Calcutta,
92 Acharya Prafulla Chandra Road, Kolkata  700 009, India
}

\date{\today}
\begin{abstract}

We show  that the  earth matter effects in the 
${\rm {\nu_e \to \nu_e}}$ 
survival probability  can be used to cleanly determine the 
third leptonic mixing angle $\theta_{13}$ and the  
sign of the atmospheric neutrino mass squared difference,
$\Delta m^2_{31}$, 
using a $\beta$-beam as a  $\nu_e$ source. 
\end{abstract}
\pacs{14.60.Pq, 14.60.Lm, 13.15.+g}
\maketitle


Determination of the third neutrino mixing angle $\theta_{13}$, 
the sign of 
$\Delta m^2_{31} \equiv
m_3^2 - m_1^2$, 
the three CP phases, and the absolute neutrino mass scale
are necessary for reconstruction 
of the neutrino mass matrix, which will have important consequences for
nuclear and particle physics, astrophysics and cosmology.  
The $\nu_e \to \nu_\mu$ transition probability
$P_{e \mu}$, 
has been identified as the ``golden channel'' \cite{golden}
for measuring the Dirac phase  
$\delta_{CP}$, 
$sgn(\Delta m^2_{31})$ and $\theta_{13}$ in 
long baseline accelerator based experiments.
However, this strength of the golden channel also brings in 
the well-known problem of 
parameter ``degeneracies'', where 
one gets multiple fake solutions in addition to the true 
one \cite{degen}. Various ways to combat this 
vexing issue have been suggested in the literature, including 
combining the golden channel with the 
``silver'' ($P_{e\tau}$) \cite{silver} 
and ``platinum'' ($P_{\mu e}$) channels. While each of them 
would have fake solutions, their combination helps
in beating the degeneracies since each channel depends 
differently on $\delta_{CP}$, 
$sgn(\Delta m^2_{31})$ and $\theta_{13}$. In this letter, 
we propose using the 
$\nu_e \to \nu_e$ survival channel, $P_{ee}$, 
which is {\it independent} of $\delta_{CP}$ and 
the mixing angle $\theta_{23}$. It is therefore {\it completely} 
absolved of degeneracies 
and hence provides a clean laboratory for the measurement 
of $sgn(\ma)$ and $\theta_{13}$. This 
gives it an edge over the 
conversion channels, which are infested with degenerate solutions.

The $P_{ee}$ survival channel has been extensively 
considered for measuring $\theta_{13}$ 
with $\anue$ produced in 
nuclear reactors \cite{white} 
and with detectors placed at a distance $\simeq 1$ km. 
Reducing
systematic uncertainties to the sub-percent level
is a prerequisite for this program and enormous 
R\&D is underway for this extremely challenging job. 
For accelerator based experiments, 
the survival channel, $P_{ee}$, has been discussed with
sub-GeV neutrinos  from a \bb{} source at CERN 
and a megaton water detector in Fr\'{e}jus at a 
baseline of 130 km \cite{donini-disapp,betapemu}. 
However, 
no significant improvement on the $\theta_{13}$ 
limit was found in \cite{donini-disapp} for a systematic 
error of $\gsim$ 5\%.
This stems mainly from the fact that
in these experiments 
one is trying to differentiate between two scenarios, 
both of which predict a large number of events,   
differing from each other by a small number  
due to the small value of $\theta_{13}$.
Also, since $sgn(\ma)$, 
is ascertained using earth matter effects, 
there is no 
hierarchy sensitivity in these survival channel 
experiments due to the 
the short baselines involved.

In this letter, we emphasize on 
the existence of large matter 
effects in the survival channel, $P_{ee}$, for an experiment with a
very long baseline. 
Recalling that
$P_{ee}= 1 - P_{e \mu} - P_{e \tau}$ and
since
for a given $sgn(\ma)$ {\it both}
$P_{e \mu}$ and $P_{e \tau}$
will either increase or decrease in matter,
the change  in $P_{ee}$ is almost twice that
in either of these channels.
Using the multi-GeV $\nue$ flux from a \bb{} source, 
we show that this large matter effect 
allows for 
significant, even maximal, deviation of $P_{ee}$ 
from unity. This, can thus be a convenient tool
to explore $\theta_{13}$.
This is in contrast to the reactor option or 
the \bb{} experimental set-up in \cite{donini-disapp}, 
where increasing the 
neutrino flux and reducing the systematic uncertainties 
are the only ways of getting any improvement on 
the current $\theta_{13}$ limit. We further show, 
for the first time, that  very good sensitivity to the 
neutrino mass ordering can also be achieved 
in the $P_{ee}$ survival channel owing to the large 
matter effects. 
We discuss plausible
experimental set-ups with the survival channel and show 
how  the 
large matter effect propels this channel, 
transforming it into a 
very
useful tool to probe
$sgn(\ma)$
and  $\theta_{13}$  
even with relatively large room for systematic uncertainties.

For simplicity,  
we start with   
one mass scale dominance (OMSD)
and the 
constant density approximation. 
OMSD implies setting $\Delta m^2_{21}=0$. 
Under these conditions,  
for neutrinos of energy $E$ and
traveling through a distance $L$, 
\be
{{P_{e e}}} =
1 - \sin^2 2 \theta_{13}^{m}
\sin^2 \left[1.27 (\Delta m^2_{31})^{m} 
{L}/{E} \right],
\label{eq:peemat}
\ee
where the mass squared difference  and mixing angle in matter are 
respectively 
\bea
{{
{(\Delta m^2_{31})^m} }} &=&
{{
\sqrt{(\Delta m^2_{31} \cos 2 \theta_{13} - A)^2 +
(\Delta m^2_{31} \sin 2 \theta_{13})^2} }}\nonumber \\
{{\sin 2 \theta^m_{13} }}
&=& {{\sin 2 \theta_{13} ~\Delta m^2_{31}/
(\Delta m^2_{31})^m }}
\label{eq:dm31}
\eea
and ${{A=2\sqrt{2}G_Fn_eE}}$
originates from the matter potential \cite{msw}. 
Here, $n_e$ is the ambient electron density. 
From Eq. (\ref{eq:peemat}), 
the largest deviation of $P_{ee}$ from unity is  
obtained when the conditions
({i})
${{\sin^2 2 \theta^m_{13} = 1}}$ and  
({ii}) $\sin^2 \left[1.27 (\Delta m^2_{31})^m L/E \right]=1$
are satisfied simultaneously.
The first condition is achieved at resonance 
which is obtained for 
${{A = \Delta m^2_{31} \cos 2 \theta_{13}}}$.
This defines the resonance energy as, 
$
E_{res} ={\Delta m^2_{31} \cos 2 \theta_{13}}/2\sqrt{2}G_Fn_e.
$
The second condition gives the energy where the $(\ma)^m$ 
driven oscillatory term is maximal,
\be
{{
E^{m}_{max} = \frac{1.27 (\Delta m^2_{31})^m L}{(2p+1) \pi/2}
}},\, p=0,1,2..
\ee
Maximum  matter effect  is  obtained  when 
${{E_{res} = E^{{m}}_{{max}}}}$ \cite{gandhiprl}, which gives,  
\be
{{
(\rho L)^{max} }} =  
{{
\frac{(2p+1) \pi \; 5.18 \times 10^3} { \tan 2\theta_{13}} ~{\rm km ~gm/cc} }}.
\label{eq:eecondtn}
\ee
$\rho$ is the matter density in gm/cc. 
This is 
the distance where $P_{ee}\simeq 0$.
Although both $E_{res}$ and $E^{m}_{max}$ 
depend on the value of $\Delta m^2_{31}$, the distance at which we 
get the maximum matter effect is 
independent of $\Delta m^2_{31}$. However, 
it is controlled by $\theta_{13}$ 
very sensitively. 
For average earth matter densities obtained using the 
Preliminary Reference Earth Model (PREM), one 
can find the typical distances at which the above conditions are 
satisfied  for various values of $\sin^2 2 \theta_{13}$ \cite{gandhi2}.
For instance for $p=0$ and 
$\sin^2 2 \theta_{13}$ = 0.2 and 0.1, these distances are 7600 km and 
10200 km respectively. For higher values of $p$ the distance 
exceeds the earth's diameter for 
$\theta_{13}$ in the current allowed range. 
Using $(\rho L)^{max}$ corresponding to the PREM  
profile, from Eq. (\ref{eq:eecondtn}) 
one can estimate that the  
condition
of maximal matter effects inside the earth's mantle
is satisfied only for $\stch \gsim 0.09$.

Under OMSD, the matter conversion probabilites are,  
$
{{P_{e x}}} =
Y_{23} \, \sin^2 2 \theta_{13}^{m}
\sin^2 \left[1.27 (\Delta m^2_{31})^{m} {L}/{E} \right],
$
where $Y_{23}=\sin^2 \theta_{23}$ for $x=\mu$ and 
 $Y_{23}=\cos^2 \theta_{23}$ for $x=\tau$. 
The maximum matter effect condition in the conversion channels 
is also given by 
Eq. (\ref{eq:eecondtn}). 
However,  
there are suppression factors,
$\sin^2\theta_{23}$
for \pemu and $\cos^2\theta_{23}$ for \petau, not present in
$P_{ee}$.  
Moreover, since $P_{ee}$ does not contain $\theta_{23}$, the 
octant ambiguity as well as 
parameter correlations due to 
uncertainty in 
$\theta_{23}$ are absent.
In addition, as mentioned earlier, the $P_{ee}$ channel does not contain the 
CP phase, $\delta_{CP}$.  
Both of these remain true  in the presence of 
non-zero $\Delta m^2_{21}$ \cite{akh}.
In our numerical work,  we solve the 
full three flavour
neutrino propagation equation assuming the PREM \cite{prem}
profile and 
keep $\Delta m^2_{21}$ and $\sin^2\theta_{12}$ fixed at their present 
best-fit values of 
$8.0 \times 10^{-5}$ eV$^2$ 
and 0.31 respectively \cite{global}. We 
assume the true value of $|\ma|=2.5\times 10^{-3}$ eV$^2$.

\begin{figure}[t]
\includegraphics[scale=.5]{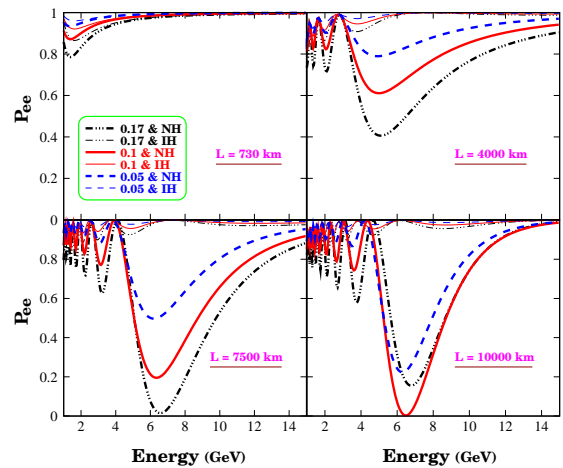}
\caption{\label{fig:fig1}
\pee in matter plotted versus neutrino energy.
Thick (thin) lines are for normal (inverted) 
hierarchy. 
}
\end{figure}
\begin{figure}[t]
\includegraphics[height=5.5cm,width=7.5cm]{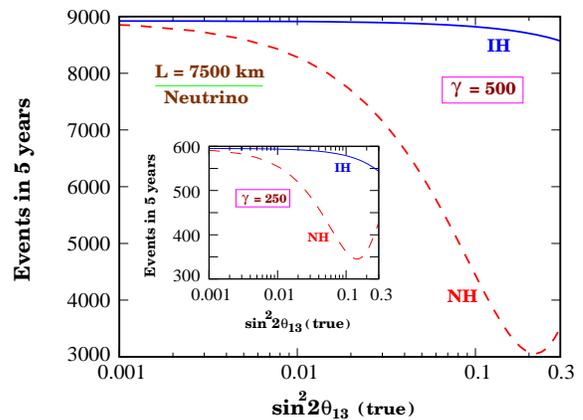}
\caption{\label{fig:fig2}
Events in 5 years vs.  
$\stch$ for the normal (dashed line) and inverted (solid line) 
hierarchy for $L=7500$ km and $\gamma=500$. The inset shows the 
same but for $\gamma=250$. 
}
\end{figure}

In Fig. \ref{fig:fig1} we plot \pee as a function of energy,
at four different $L$ 
and for three values of $\sin^2 2\theta_{13}$. 
The plots confirm that maximal matter effects 
come at $L\simeq 10000$ km and $L\simeq 7500$ km for 
$\stch=0.1$ and 0.17 respectively for the normal hierarchy (NH). 
For the inverted hierarchy (IH) there is no significant matter effect 
for $\nu_e$.   
This large difference in the 
probabilities for NH and IH
can be exploited for the determination of  
$sgn(\ma)$.
Further, since the matter effect is a sensitive function of 
$\theta_{13}$ it may also be possible to obtain information on 
this angle. We can also see that for a given value of $\stch$ 
($\gsim 0.09$)
and $E$, the matter effect increases (almost linearly) with $L$, 
until the $L$ for maximal matter effect  
is reached, beyond which matter effect falls. 
For values of $\sin^2 2\theta_{13} <  0.09$ the condition 
for maximum matter effect 
is not met inside the earth's mantle  
and hence the matter  effect and 
sensitivity 
to both hierarchy as well as $\theta_{13}$ 
increase with $L$.

In what follows, we will show how, in a plausible experiment, one 
can use this near-resonant matter 
effect in the survival channel, $P_{ee}$,   
to constrain $\theta_{13}$ and $sgn(\ma)$.
Fig. \ref{fig:fig1} shows that 
the requirements for such a program include 
a $\nu_e$ beam, baselines of at least a few thousand km and 
average energies around 6 GeV. The detector 
should be able to observe $e^-$ 
unambiguously at these energies. 

Pure $\nu_e/\bar{\nu}_e$ beams can be produced from 
completely ionized radioactive ions 
accelerated to high energy decaying through
the beta process
in a storage ring, popularly known as
$\beta$-beams \cite{zucc,betapemu,sanjib1}.
The ions considered 
as possible sources for beta beams are $^{18}{Ne}$ and $^{8}{B}$ 
for $\nu_e$ and $^{6}{He}$ and $^{8}{Li}$ for $\anue$.  
The end point energies of  
$^{6}{He}$ and $^{18}{Ne}$ are $\sim 3.5$ MeV 
while for $^{8}{B}$ and $^{8}{Li}$ this can be larger $\sim$ 13-14 MeV 
\cite{rubbia}.
For the Lorentz boost factor $\gamma=250(500)$  the $^{6}{He}$ and
$^{18}{Ne}$ sources
have peak energy around $\sim 1(2)$ GeV whereas for the 
$^{8}{B}$ and $^{8}{Li}$ sources the peak occurs at a higher value 
around $\sim 4(7)$ GeV. Since the latter is in the ball-park of
the energy necessary for near-resonant matter effects 
as discussed above, we will work with 
$^8B$ ($^8Li$) as the source ion for the $\nue$ ($\anue$) 
$\beta$-beam and $\gamma=250$ and 500. 
A CERN $\beta$-beam facility with 
$\gamma\simeq 250$ 
should be possible with the existing SPS, while $\gamma \leq 500$ 
could be achieved with upgrades of the existing accelerators.
The Tevatron is also being projected as a plausible 
accelerator for the $\beta$-beam.  

Water \chr detectors have excellent 
capability of separating electron 
from muon events. Since this technology is 
very well known, megaton water detectors are considered to be 
ideal for observing $\beta$-beams. 
Such detectors do not have any charge identification capacity. 
But in a $\beta$-beam, the   
$\beta^{-}$ and $\beta^{+}$ emitters can be stacked in different bunches
and the timing information at the detector can  help to identify   
the $e^{-}$ and $e^{+}$ events \cite{tnova}.  

It is well known that there are no
beam induced backgrounds for $\beta$-beams.
In this experimental set-up,  the process 
$\nu_e \to \nu_\tau \to \tau^- \to e^-$ could mimic the signal.
We have checked that the background to signal ratio for these events in 
the relevant energy range is $\sim 10^{-2}$ and 
can be 
neglected for the disappearance mode. 
$e^-$ events from $K$ and $\pi^-$ decays are also
negligible.
The atmospheric background can be estimated in the beam off mode
and reduced through directional, timing, and energy cuts.

Proposals for megaton water detectors include 
UNO \cite{uno} in USA, HyperKamiokande \cite{hk} in Japan 
and MEMPHYS \cite{memp} in Europe. If the \bb{} is produced at 
CERN, then baselines in the range 7000-8600 km would be 
possible at any of the proposed locations for the UNO
detector. Likewise, if the \bb{} source be at FNAL, then the 
far detector MEMPHYS would allow for $L=7313$ km.  
HyperKamiokande could also be considered as the far detector and 
in that case  $L=10184$ ($9647$) km if the source be at 
FNAL (CERN).

For our numerical analysis 
we use the standard $\chi^2$ technique with
$\chi^2_{total} = \chi^2_{pull} + \chi^2_{prior}$,
where $\chi^2_{prior}=\left [(|\Delta m^2_{31}|-
|\mat|)/\sigma(\Delta m^2_{31}) \right ]^2$. 
In $\chi^2_{pull}$ we  
consider 2\% $\beta$-beam flux
normalisation error and 2\% error for detector systematics 
which is more conservative 
than the value of 2\% usually considered in current 
literature \cite{donini-disapp,betapemu}. 
For the full definition of $\chi^2_{pull}$ and 
details of our statistical analysis, we refer to
\cite{bino}.
The prospective ``data'' is generated 
at the ``true'' values of oscillation parameters,
assuming 440 kton of 
fiducial volume for the detector 
with 90\% detector efficiency, threshold of 4 GeV and 
energy smearing of width 15\%. 
For the $\nue$ \bb{} we have assumed $1.1 \times 10^{18}$ 
useful $^8B$ decays per year  
and show results for 5 years of running of this beam.  
The number of events as a function of $\stch$
at $L=7500$ km
with a $\gamma=500$ $\nue$ \bb{} is shown 
in Fig. \ref{fig:fig2} for NH and 
IH. The inset in Fig. \ref{fig:fig2} 
shows the  
number of 
events in 5 years expected from a lower $\gamma=250$. 
We have used the neutrino-nucleon interaction cross-sections 
from \cite{globes}.  

\begin{figure}[t]
\includegraphics[height=5.0cm,width=7.0cm]{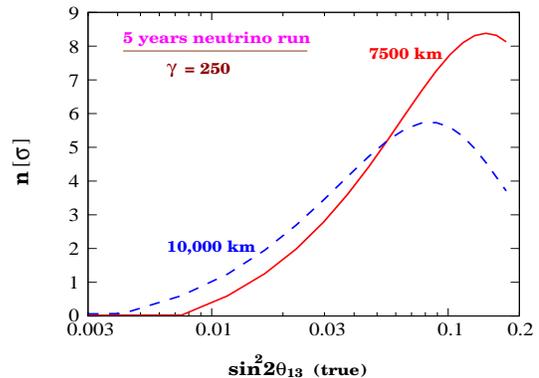}
\caption{\label{fig:fig3}
Sensitivity to hierarchy for $L=7500$ (solid line) and 10000 
km (dashed line) and $\gamma=250$,
as a function of true value of $\stch$.
}
\end{figure}

\begin{figure}[t]
\includegraphics[height=6.0cm,width=8.0cm]{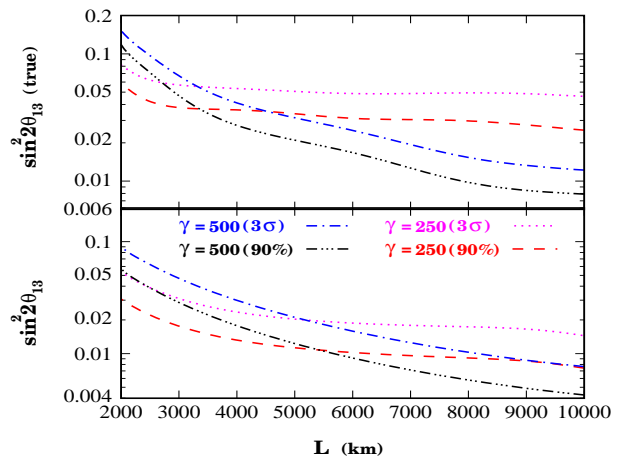}
\caption{\label{fig:fig4}
The upper panel shows  the smallest value of true
$\sin^2 2\theta_{13}$ at which normal 
and inverted hierarchy can be distinguished
while the lower panel gives the sensitivity to $\sin^2 2\theta_{13}$
at various
baselines 
at 90\% and 3$\sigma$ C.L., for two values of $\gamma$. 
}
\end{figure}

In Fig. \ref{fig:fig3} we show the sensitivity ($n \sigma, n=\sqrt{\chi^2}$)
of 
the survival channel to the neutrino mass ordering
for $L=7500$ and 
10000 km and $\gamma=250$. 
If the true value of $\stch=0.05$, 
then one can rule out the inverted hierarchy at the 4.8$\sigma$ 
(5.0$\sigma$) level with $L=7500$ (10000) km. 
For $L=7500$ (10000) km, 
the wrong inverted hierarchy can be disfavored at the 
90\% C.L. 
if the true value of $\stch>0.03$ (0.025).
The sensitivity improves significantly if we 
use $\gamma=500$ instead of 250, since (i) the flux 
at the detector increases, and (ii) the flux peaks at $E$ closer 
to 6 GeV, where we expect largest matter effects \cite{bino}. 
For  $\gamma=500$, the inverted hierarchy can be disfavored at
2.6$\sigma$(3.8$\sigma$)  
for a lower value of $\sin^2 2\theta_{13} = 0.015$.
Minimum values of $\stch$ 
at which the inverted hierarchy can 
be ruled out 
at 90\% and 3$\sigma$ C.L. for different values of $L$ 
are shown in  the upper panel of Fig. \ref{fig:fig4} for $\gamma=250$ 
and 500. 
From the figure one can see that for 
$\gamma=500$ and $L$ = 7500 (10000) km
the wrong inverted hierarchy can be disfavored at the 
90\% C.L. if the true value of $\stch>1.0\times 10^{-2}$ 
($8.0\times 10^{-3}$). 
If instead we use a total systematic error of 5\% 
then we get the above sensitivity limits as 
$\stch>1.6\times 10^{-2}$
($1.2\times 10^{-2}$) at 90\% C.L.


If the true value of $\stch$ turns out to be smaller 
than the sensitivity reach shown in the upper panel  of 
Fig. \ref{fig:fig4} for 
a given $L$, then it would not be 
possible to determine the 
hierarchy at the given C.L. 
However, we would still be able to put better 
constraints on $\stch$ itself. 
The lower panel of Fig. \ref{fig:fig4} 
demonstrates as a function of $L$ 
the sensitivity to $\theta_{13}$, i.e.,
the minimum value of $\stch$ which can be 
statistically distinguished from $\stch=0$ 
at 90\% and 3$\sigma$ C.L. 
Both Figs. \ref{fig:fig3} and \ref{fig:fig4} show that 
the sensitivity 
improves with $L$, even though the 
 flux 
falls 
as $1/L^2$. 
This results from matter 
effects increasing with $L$, as noted before. 
For $L=7500$ (10000) km, we can constrain 
$\stch < 6.3  \times 10^{-3}$ ($4.3 \times 10^{-3}$)
at the 90\% C.L.  for $\gamma=500 $.  
 For a 5\% systematic error the above numbers are changed to 
$\stch < 1.0  \times 10^{-2}$ ($7.3 \times 10^{-3}$). 

How does this compare with alternate possibilites ? 
If the energy can be reconstructed accurately, 
then the result can be improved further.  
For instance, for $L$=7500 km, if one could
preferentially  select the energy
in the range 5.0-7.5 GeV,
then the normal and inverted hierarchy would be 
differentiated for $\sin^2 2\theta_{13}=7.47 \times 10^{-3}$  
at 90\% C.L. for $\gamma =500$.  
   
Using the $P_{e \mu}$ channel and a  $\beta$-beam source 
at a distance of 7152 km (CERN-INO) from the detector,  
the NH can be distinguished from IH at 90\% C.L. for 
$\gamma =500$ if  $\sin^2 2\theta_{13}>7.7 \times 10^{-3}$ \cite{bino}. 
For a neutrino factory at 7500 km, the wrong IH can be 
discarded at the $3\sigma$ C.L. if 
$\sin^2 2\theta_{13}>3 \times 10^{-4}$ \cite{optimnufact}.

We have presented our results using a $\nu_e$ beam and assuming 
NH to be the true hieararchy. 
Similar results can also be obtained with  a $\anue$ beam for IH.
It is also possible to run both beams simultaneously.  


In conclusion, we propound the possibility of using 
large matter effects in   
the survival channel, $P_{ee}$, at long baselines for 
determination of  the
neutrino mass ordering ($sgn(\Delta m^2_{31})$) 
and the yet unknown leptonic mixing
angle $\theta_{13}$. 
Matter effects in the transition probabilites 
$P_{e \mu}$ and $P_{e \tau}$ act in consonance 
to give an almost two-fold 
effect in the survival channel. In addition, 
the problem of spurious solutions due to the leptonic CP phase 
and the atmospheric mixing angle $\theta_{23}$ does   
not crop up. 
The development of
$\beta$-beams as sources of 
pure $\nu_e/\bar\nu_e$ beams 
enables one to exclusively study the $P_{ee}$
survival probability and adds a new direction to the prospects of a 
future $\beta$-beam. 

We thank F. Terranova for a useful communication.


\end{document}